\documentclass[twoside,11pt]{article}

\usepackage{jmlr2e}

\usepackage{hyperref}       
\usepackage{url}            
\usepackage{booktabs}       
\usepackage{amsfonts}       
\usepackage{nicefrac}       
\usepackage{microtype}      

\usepackage{amsmath}
\usepackage{graphicx}  
\usepackage{subfigure}
\usepackage{caption}

\begin{document}
	
	\title{A Computational Model of Learning Flexible Navigation in a Maze by Layout-Conforming Replay of Place Cells}
	\author{\name Yuanxiang Gao \\
		\addr University of Electronic Science and Technology of China\\
		\email gaoyuanxiang123@alu.uestc.edu.cn}
	
	\bibliographystyle{plain}
	
	\maketitle
	
	\begin{abstract}
	Recent experimental observations have shown that the reactivation of hippocampal place cells (PC) during sleep or immobility depicts trajectories that can go around barriers and can flexibly adapt to a changing maze layout. Such layout-conforming replay sheds a light on how the activity of place cells supports the learning of flexible navigation of an animal in a dynamically changing maze. However, existing computational models of replay fall short of generating layout-conforming replay, restricting their usage to simple environments, like linear tracks or open fields. In this paper, we propose a computational model that generates layout-conforming replay and explains how such replay drives the learning of flexible navigation in a maze. First, we propose a Hebbian-like rule to learn the inter-PC synaptic strength during exploring a maze. Then we use a continuous attractor network (CAN) with feedback inhibition to model the interaction among place cells and hippocampal interneurons. The activity bump of place cells drifts along a path in the maze, which models layout-conforming replay. During replay in rest, the synaptic strengths from place cells to striatal medium spiny neurons (MSN) are learned by a novel dopamine-modulated three-factor rule to store place-reward associations. During goal-directed navigation, the CAN periodically generates replay trajectories from the animal's location for path planning, and the trajectory leading to a maximal MSN activity is followed by the animal. We have implemented our model into a high-fidelity virtual rat in the MuJoCo physics simulator. Extensive experiments have demonstrated that its superior flexibility during navigation in a maze is due to a continuous re-learning of inter-PC and PC-MSN synaptic strength.
	 
	\end{abstract}
	\raggedbottom
	
	\section{Introduction}
	It has been observed that, during sleep or wakeful immobility, hippocampal place cells (PC) spontaneously and sequentially fire, similar to the pattern of activity during movement periods \cite{Lee2002, Foster2006, Pfeiffer2013, Stella2019}. Conventionally, studies of such hippocampal replay were restricted to animals moving in simple environments like linear tracks or open fields. Recently, an interesting study \cite{Widloski2022} observed that, for a rat moving in a reconfigurable maze, the replay trajectories conform to the spatial layout and can flexibly adapt to a new layout. Such layout-conforming replay is the key to understanding how the activity of place cells supports the learning of flexible navigation, a long-standing open question in computational neuroscience. 
	
	Existing models \cite{Azizi2013, Itskov2011, Hopfield2009, Romani2015} of hippocampal replay use a continuous attractor network (CAN) with spike-frequency adaptation or short-term synaptic depression \cite{Romani2015} to generate replay of place cells. In these models, the synaptic strength between a pair of place cells is pre-configured by a decaying function of the Euclidean distance between their place field centers. In a maze, such Euclidean synaptic strength introduces a strong synaptic coupling between a pair of place cells with firing field centers located at two sides of a thin wall. Accordingly, the activity bump of place cells will pass through the wall without conforming to the layout. Besides, these models lack the modeling of how the replay activity drives downstream circuits, such as striatum, to perform high-level functional roles, such as reward-based learning, planning and navigation. Although reinforcement learning (RL) algorithms \cite{Sutton1998} can serve as candidate models of these functional roles \cite{Foster2000, Johnson2005, Russek2017, Gustafson2011, Brown1995}, these algorithms are designed to solve much more general and abstract control problems in engineering domains, falling short of biological plausibility.
	
	In this paper, we propose a computational model for generating layout-conforming replay and explaining how such replay supports the learning of flexible navigation in a maze. First, we model the firing field of a place cell by a function decaying with the shortest path distance from its field center, which conforms to experimental observations \cite{Skaggs1998, Widloski2022, Gustafson2011}. Then we propose a Hebbian-like rule to learn the inter-PC synaptic strength during exploration. We find that the synaptic strength between a pair of place cells encodes the spatial correlation of their place fields and decays with the shortest path distance between their field centers. With learned inter-PC synaptic strength, the interactions among place cells are modeled by a CAN with feedback inhibition from hippocampal GABAergic interneurons \cite{Stark2014, Schlingloff2014}. The drift of the activity bump of place cells under vanishing or non-zero visual inputs models layout-conforming replay during rest and goal-directed navigation, respectively.
	
	During replay in rest, the synaptic strength from place cells to medium spiny neurons (MSN) in striatum is learned by a novel three-factor learning rule based on a replacing trace rule \cite{Singh1996, Seijen2014}, rather than the conventional accumulating trace rule \cite{Sutton2018, Gerstners2018, Izhikevich2007}. This learning rule strengthens a PC-MSN synapse proportional to the co-firing trace of this pair of PC and MSN multiplied by the dopamine release at the synapse \cite{Yagishita2014, Kasai2021}. After replay, the PC-MSN synaptic strength encodes the geodesic proximity between the firing field center of each place cell and the goal location. As a result, the activity of the MSN population will ramp up when an animal approaches the goal location, which conforms to a line of experimental observations \cite{Meer2011, ONeal2022, Atallah2014, Sjulson2018, London2018, Howe2013}. During goal-directed navigation, the attractor network periodically generates a series of replay trajectories from the rat’s location to lookahead along each explorable path \cite{Johnson2007, Pfeiffer2013}, and the trajectory leading to a maximal MSN activity is followed by the rat with a maximal probability.
	
	We have implemented our model into a high-fidelity virtual rat that reproduces the average anatomical features of seven Long-Evans rats in the MuJoCo physics simulator \cite{Merel2020, Todorov2012}. Extensive experiments have demonstrated that the virtual rat shows superior flexibility during navigation in a dynamically changing maze. We have observed that, after the layout changes, the inter-PC synaptic strength is updated to re-encode the adjacency relation between locations in the new layout. As a result, the replay trajectories adapt to the new layout so that after replay the MSN activity will ramp up along new paths to the goal location. Periodical planning with respect to the updated MSN activity explains the navigational flexibility of the virtual rat.
	
	\section{Model}
	\subsection{Learning inter-PC synaptic strength}
	
	For a population of $M$ place cells, the firing rate of the $i$-th place cell is denoted by $r_i$. Each place cell has a preferential location, denoted by $\mathbf{x}^{(i)}$, where its firing rate is maximal among all locations. The preferential location of each place cell is uniformly arranged on a regular grid of a square maze without considering grid points that fall into areas under walls. Let $J_{ij}$ denote the strength of the synapse from cell $j$ to cell $i$. Before learning $J_{ij}$, the firing rate of each place cell is modeled by a function of the rat's location $\mathbf{x}$ given by,  
	\begin{equation}
		r_i(\mathbf{x}) = \exp(-{\rm D}(\mathbf{x}^{(i)}, \mathbf{x})/\sigma), \forall i, 
		\label{geoField}
	\end{equation}
	where $\sigma$ is a scale parameter and ${\rm D}(\mathbf{x}^{(i)}, \mathbf{x})$ is the shortest path distance between $\mathbf{x}^{(i)}$ and $\mathbf{x}$. The firing field in Equation~(\ref{geoField}) is a geodesic place field \cite{Gustafson2011}, which conforms to the constraints imposed by walls and is a more accurate model of firing fields observed in a maze \cite{Skaggs1998, Widloski2022, Gustafson2011} compared to Gaussian place fields observed in open fields \cite{Keefe1996, Foster2000}.
	
	For a rat randomly exploring a maze with many walls, a pair of place cells with a small shortest path distance between their preferential locations will often fire in close temporal order. According to symmetric spike-timing dependent plasticity (STDP) \cite{Mishra2016, Isaac2009}, the pair of recurrent synapses between this pair of place cells will become stonger than pairs of place cells with larger shortest path distance between their preferential locations. Accordingly, for a sequence of positions $\mathbf{X} = \{\mathbf{x}_0, \mathbf{x}_1, ..., \mathbf{x}_N\}$ visited by a virtual rat during random exploration, the synaptic strength matrix is symmetrically updated by the following Hebbian-like rule, 
	\begin{equation}
		\mathbf{J}^{(n+1)} = \mathbf{J}^{(n)} + \alpha_1 (\mathbf{r}(\mathbf{x}_n)^{\rm T} \mathbf{r}(\mathbf{x}_{n}) - \mathbf{J}^{(n)}),
		\label{Hebb}
	\end{equation}
	where $\mathbf{r}(\mathbf{x}_n) = \{r_i(\mathbf{x}_n), i = 1,...,M\}$ is the row vector of place cell firing rates at position $\mathbf{x}_n$,  $\alpha_1$ is the learning rate, and $\mathbf{J}^{(0)}$ is initialized as a zero matrix. The conventional Hebbian learning rule \cite{Muller1996a, Stringer2002, Haykin1998} increments inter-PC synaptic strength proportional to the product of firing rates of a pair of place cells hence inter-PC synaptic strength will increase without bound and fail to converge. The proposed Hebbian-like rule is free from such convergence issues.
	
	The learning rule in Equation~(\ref{Hebb}) is a stochastic approximation process \cite{Nevelson1976} that solves the following equation as $n \to \infty$,
	\begin{equation}
		J_{ij}^* = \mathbb{E}[r_i(\mathbf{x}) r_j(\mathbf{x})], \forall i, j, 
		\label{corr}
	\end{equation}
	where the expectation is taken over possible locations during exploration. As shown by Equation~(\ref{corr}), the synaptic strength between a pair of place cells encodes the correlation of their firing fields, which characterizes the geodesic proximity of their field centers. 
	
	\subsection{Hippocampal replay by a continuous attractor network}
	After learning the synaptic strength matrix, the activity of place cells can change independently with respect to the rat's current location $\mathbf{x}$, due to synaptic interactions among place cells. Thus, $r_i$ is also a function of time, denoted by $r_i(\mathbf{x}, t)$. Since $\mathbf{x}$ is fixed during replay, $r_i(\mathbf{x}, t)$ is written as $r_i(t)$ for simplicity. $r_i(t)$ obeys following differential equations,
	\begin{equation}
		\left\{ 
		\begin{array}{lc}
			\tau_r \dot{r}_i= - r_i +\left[\sum\limits_{j \neq i} J_{ij} r_j + E_i - I_i - h_0\right]^+, \\
			\\
			\tau_I \dot{I}_i = - I_i + c_I r_i, 
		\end{array}
		\right.
		\label{CANN}
	\end{equation}
	where $\tau_r$ and $\tau_I$ are time constants, $h_0$ is a threshold, $E_i$ is the external input to cell $i$, $I_i$ is the feedback inhibition due to the reciprocal connection between cell $i$ and hippocampal GABAergic interneurons \cite{Pelkey2017}, $c_I$ is the strength of feedback inhibition, and $[x]^+ = \max\{x, 0\}$ is the neural transfer function given by a threshold-linear function. The feedback inhibition in Equation~(\ref{CANN}) is supported by observations that hippocampal sharp wave ripples (SWRs) containing replay events are generated by feedback inhibition from parvalbumin-containing (PV+) basket neurons \cite{Schlingloff2014, Stark2014, Buzsaki2015}.
	
	In Equation~(\ref{CANN}), the external input $E_i$ characterizes the association of place cell $i$ with visual cues at the rat's location. $E_i$ is typically a decaying function of the distance between the rat's location and the firing field center of place cell $i$ \cite{Stringer2002, Hopfield2009, Stringer2004}. Accordingly, $E_i$ is given by,
	\begin{equation}
		E_i(\mathbf{x}) = \mathcal{A} \cdot \exp(-{\rm D}(\mathbf{x}^{(i)}, \mathbf{x})/\sigma), \forall i, 
		\label{firingField}
	\end{equation}
	where $\mathcal{A}$ is the amplitude of the external input. Under Euclidean synaptic strength without feedback inhibition (e.g., $c_I = 0, I_i \equiv 0$), the steady state solution of $\mathbf{r}$, mapped to each preferential location, is a localized Gaussian-like bump centered at the rat's location in an open field \cite{Fung2010, Wu2008}, due to locally strong excitatory coupling among place cells with firing fields nearby the rat’s location. The learned synaptic strength in Equation~(\ref{CANN}) has similar locally strong coupling so that its steady state solution without feedback inhibition is also a localized Gaussian-like bump centered at the rat's location in the maze.
	
	With both external input and feedback inhibition, the activity bump may move away from the rat's location and its mobility modes are determined by both factors in a competitive way. The external input tends to stabilize the bump at $\mathbf{x}$ and the feedback inhibition tends to deviate the bump from its current location. Under a fixed strength of feedback inhibition, if the external input is strong enough, the feedback inhibition will fail to deviate the stable activity bump. In contrast, if the external input vanishes, the activity bump will freely drift along paths in the maze, resembling replay during sleep or rest without sensory drive \cite{Stella2019}. Most interestingly, for an external input with intermediate amplitude, the drift of the activity bump will periodically jump back to the rat's location to regenerate a drift trajectory along other paths, resembling replay during goal-directed navigation \cite{Johnson2007, Pfeiffer2013, Widloski2022}.
	
	\subsection{Learning PC-MSN synaptic strength}
	During replay in sleep or brief rest, the synaptic strengths from place cells to a population of medium spiny neurons (MSN) in the striatum are selectively strengthened to store place-reward associations \cite{Sjulson2018, Sosa2020, Lansink2009, Trouche2019}. The learning of PC-MSN synaptic strength is achieved by dopamine-modulated synaptic plasticity of PC-MSN synapses  \cite{Floresco2001, Jay2003, Brzosko2019}. Precisely, the co-firing of a place cell and MSN leaves a chemical trace (i.e., Ca$^{2+}$ influx) at the synapse (connection) from the place cell to the population of MSN \cite{Yagishita2014, Kasai2021}. After co-firing, the Ca$^{2+}$ is gradually taken up by organelles hence the chemical trace decays exponentially at the synapse \cite{Yagishita2014, Abrams1988}. Before the Ca$^{2+}$ is fully absorbed,  if the dopamine concentration increases/decreases from a base level at the synapse, the synapse will be strengthened/depressed proportional to both the chemical trace and the increase/decrease of the dopamine concentration \cite{Kasai2021, Yagishita2014, Gerstners2018}. These experimental observations are modeled as follows.
	
	Let $W_i(t)$ denote the synaptic strength from place cell $i$ to the MSN population at time $t$. The activity of the MSN population at time $t$ is modeled by \cite{Sjulson2018},
	\begin{equation}
		V(t) = \sum\limits_i W_i(t) r_i(t).
	\end{equation}
	One recent study \cite{Gauthier2018} has found a population of goal cells in hippocampus with their place field centers always tracking the goal location $\mathbf{x}_{g}$. Let $G(t)$ denote the activity of the population of goal cells, and $U_i$ denote the synaptic strength from place cell $i$ to the goal cell population. Accordingly, the activity of the goal cell population is modeled by,
	\begin{equation}
		G(t) = \sum\limits_i U_i r_i(t),
		\label{goalCell}
	\end{equation}
	where $U_i = \exp(-{\rm D}(\mathbf{x}^{(i)}, \mathbf{x}_g)/\xi)$, characterizing the stronger connection between place cells with firing fields nearby the goal location and the goal cell population. Let $z_i(t)$ denote the chemical trace at the synapse from place cell $i$ to the MSN population at time $t$. Let $\delta(t)$ denote the deviation of dopamine concentration upon a base level at PC-MSN synapses at time $t$.
	
	During replay in sleep or rest, the PC-MSN synaptic strengths are updated by the following three-factor rule,
	\begin{equation}
		\left\{ 
		\begin{array}{lc}
			\dot{W}_i(t)= \alpha_2 z_i(t) \delta(t), \\
			\\
			z_i(t) = r_i(t) V(t) I_{[r_i(t) V(t) > q]}, \\
			\\
			\dot{z}_i(t) = - \frac{z_i(t)}{\tau_z} I_{[r_i(t) V(t) \leq q]},\\
			\\
			\delta(t) = G(t) + \dot{V}(t),
		\end{array}
		\right.
		\label{neoHeb}
	\end{equation}
	where $\alpha_2$ is the learning rate, $W_i(0)$ is initialized to be zero, $q$ is a small threshold for judging whether PC $i$ and MSN are co-firing, and $I_{[\cdot]}$ is the indicator function that equals 1 if the condition holds and 0 otherwise. $\tau_z$ is a time constant. In Equation~(\ref{neoHeb}), the dynamic of $z_i(t)$ characterizes that the concentration of Ca$^{2+}$ at a PC-MSN synapse encodes the instantaneous joint firing rate of PC $i$ and MSN when PC $i$ and MSN are co-firing or decays exponentially otherwise \cite{Helmchen1996, Wang1998}. Computationally, this novel trace update rule is a continuous-time generalization of the replacing trace rule \cite{Singh1996, Seijen2014} (with $q=0$ without postsynaptic factors), which has better learning efficiency than the conventional accumulating trace rule \cite{Sutton2018, Gerstners2018, Izhikevich2007}. In Equation~(\ref{neoHeb}), $\delta(t)$ characterizes the activity of VTA dopamine neurons, which is a summation of the signal on a disinhibitory pathway from the goal cell population \cite{Luo2011} and the time derivative signal of MSN activity \cite{Kim2020}. The time derivative signal of MSN activity is the summation of the signal on a disinhibitory direct pathway and the signal on an inhibitory indirect pathway from the MSN population \cite{Kim2020, Keiflin2015, Morita2012}. $\delta(t)$ is a continuous-time generalization of the temporal-difference signal \cite{Schultz1997, Uchida2017}.
	
	\subsection{Goal-directed navigation by path planning}
	During replay in goal-directed navigation, the population vector of place cells $\mathbf{p}(t) = \sum_{i} r_i(t) \mathbf{x}^{(i)}$ approximately tracks the center of the activity bump hence $\mathbf{p}(t)$ depicts a discontinuous trajectory in the maze. Let $t_0^{(i)}$ denote the instant of the $i$-th time such that $\mathbf{p}(t)$ leaves the circular region with radius $d$ from the rat's location $\mathbf{x}$, and $t_1^{(i)}$ denote the instant of the $i$-th time such that $\mathbf{p}(t)$ jumps back into the same circular region. $\{\mathbf{p}(t),\  t_0^{(i)}<t<t_1^{(i)}\}$ defines the $i$-th sub-trajectory with the initial direction given by $\mathbf{n}^{(i)}=\mathbf{p}(t_0^{(i)}) - \mathbf{x}$. 
	
	A line of studies \cite{Meer2011, ONeal2022, Atallah2014, Sjulson2018, London2018, Howe2013} has shown that the activity of MSN will ramp up when a rat is approaching the goal location. Previous studies \cite{Widloski2022, Pfeiffer2013, Xu2019} have also observed that rats show tendencies to follow replay trajectories that are approaching the goal location. Accordingly, the maximal MSN activity during the generation of the $i$-th sub-trajectory serves as the motivation for actually moving along $\mathbf{n}^{(i)}$, which is consistent with observations that the activity of MSN facilitates locomotion \cite{Kravitz2010, Freeze2013}. 
	
	Formally, if $K$ sub-trajectories are generated, the probability of choosing the $k$-th direction $\mathbf{n}^{(k)}$ to move is given by, 
	\begin{equation}
		P_{\mathbf{x}}(\mathbf{n}^{(k)}) = \frac{{\rm exp}(\beta \max_{t\in[t_0^{(k)},\  t_1^{(k)}]} V(t))}{\sum\limits_{i=1}^{K} {\rm exp}(\beta \max_{t\in[t_0^{(i)},\  t_1^{(i)}]} V(t))},
		\label{behavPol}
	\end{equation}
	where $\beta$ characterizes the greediness level of behavior. The locomotion of the virtual rat is divided into periods of movements. During each movement period, the virtual rat first performs a period of replay to evaluate explorable paths, and then samples a direction from Equation~(\ref{behavPol}) to move until the end of this movement period.
	
	\section{Experiments}
	\subsection{Experimental setup}
\begin{figure*}[h]
		\subfigure[\label{setup1}]{\includegraphics[width=1.3in, height=1.1in]{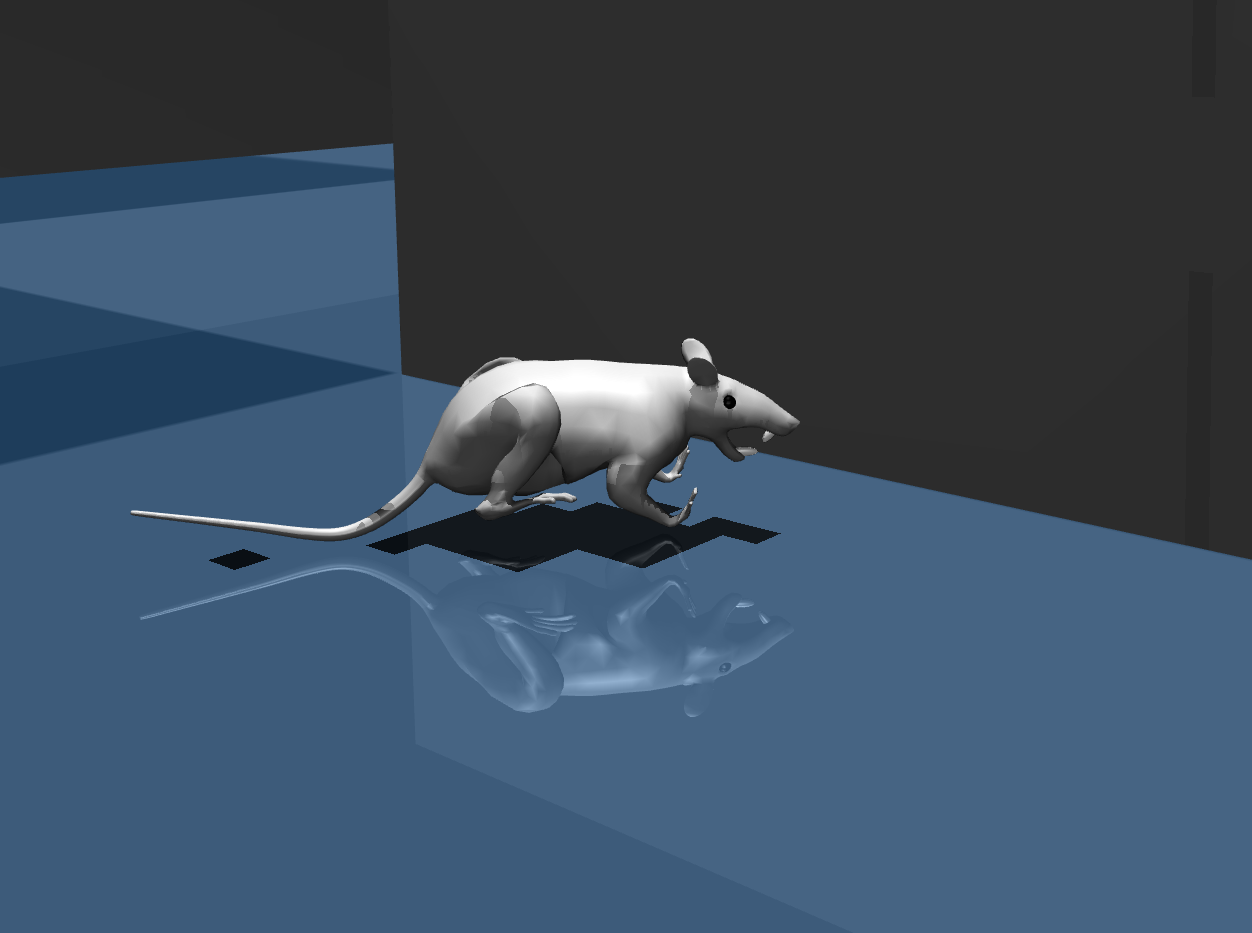}}
		\hfil
		\subfigure[\label{setup2}]{\includegraphics[width=1.1in, height=1.1in]{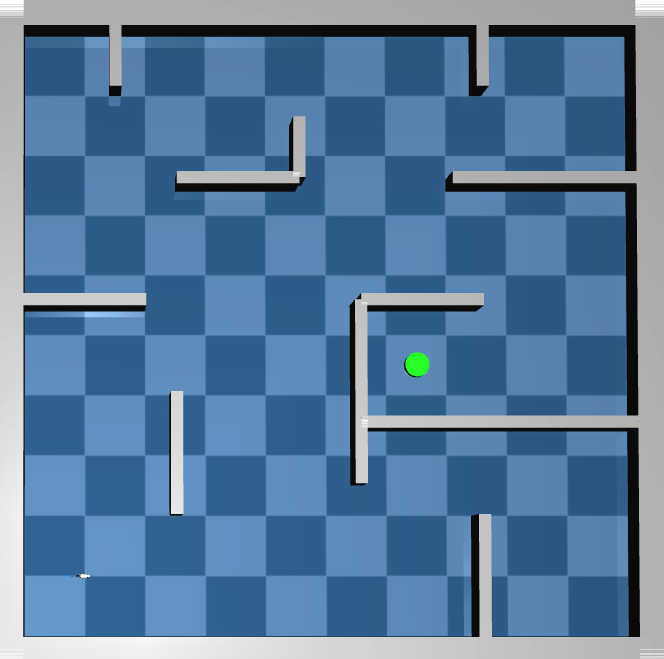}}
		\hfil
		\subfigure[\label{setup3}]{\includegraphics[width=1.1in, height=1.1in]{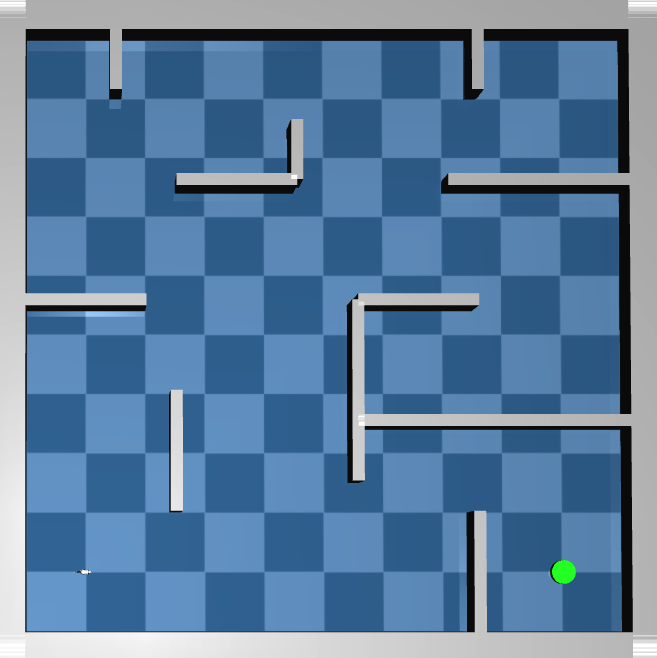}}
		\hfil
		\subfigure[\label{setup4}]{\includegraphics[width=1.1in, height=1.1in]{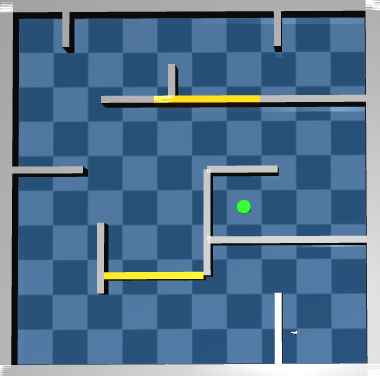}}
		\hfil
		\subfigure[\label{setup5}]{\includegraphics[width=1.1in, height=1.1in]{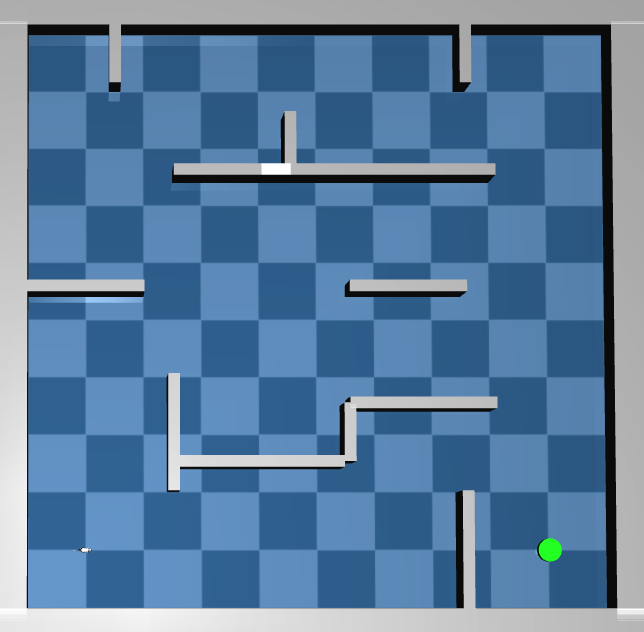}}
		\caption{The virtual rat and maze environments. (a) The virtual rat used in experiments. (b) The maze environment used in the goal-fixed experiment. (c) The maze environment used in the goal-changing experiment. (d) The maze environment used in the detour experiment. (e) The maze environment used in the shortcut experiment. }
		\label{}
	\end{figure*}

	\textbf{Subject.} The subject is a virtual rat that reproduces the anatomical features of Long-Evans rats in the MuJoCo physics simulator \cite{Merel2020, Todorov2012}, as shown in Figure \ref{setup1}. There are in total 67 bones modeled following average measurements from seven Long-Evans rats. At each simulation step, the proprioceptive input to the virtual rat is an 148-dimension vector that includes the position, velocity and angular velocity of each joint. The motor output from the virtual rat is a 38-dimension vector that contains the torque applied to each joint. Compared to numerical simulations typically used in rodent-navigation modeling literature, the behavior of the MuJoCo rat is much closer to the behavior of an actual rat. Therefore, a computational model that works on the MuJoCo rat might provide a deeper understanding about how neuronal activity relates to external behavior.
	
	\textbf{Control scheme.} Controlling the rat to run is a challenging high-dimensional continuous control task \cite{Schulman2016}, so we use the TD3 deep RL algorithm \cite{Fujimoto2018} to pretrain the rat in an open field to master several basic locomotion policies, including running forward along the current body direction, turning the body by a certain angle then running forward. The trained turning angles are 45$^\circ$, 90$^\circ$, 135$^\circ$, 180$^\circ$, -45$^\circ$, -90$^\circ$ and -135$^\circ$. Each basic locomotion policy is represented by a feedforward neural network that maps the proprioceptive input into the correct motor output to generate the desired locomotion. The pretraining details and hyperparameters used for the TD3 algorithm see Appendix \ref{pretrain} and Appendix \ref{hyperparas}, respectively. For moving along a direction sampled from Equation~(\ref{behavPol}), the rat compares its current body direction with the sampled direction and invokes the closest locomotion policy to approximately run along the sampled direction for a period. Such control scheme constitutes a hierarchical controller \cite{Merel2019} where the feedforward neural networks are low-level controllers and the CAN and MSN serve as the high-level controller.
	
	For testing the proposed computational model, we build a large scale 10×10 m maze in MuJoCo as shown in Figure \ref{setup2}. We conduct following four experiments in the maze environment.
	
	\textbf{Goal-fixed experiment.} In the goal-fixed experiment, a location in the maze is set as the goal location, as shown by the green point in Figure \ref{setup2}. The rat first randomly explores the maze for 50 trials starting from random initial positions to learn the inter-PC synaptic strength by Equation~(\ref{Hebb}). Every exploration trial contains 6000 simulation steps, corresponding to 120 s simulated time. Before every 150 simulation steps (i.e., 3 s), the rat randomly chooses a turning angle and runs along this direction in next 150 simulation steps. After exploration trials, the rat enters a brief rest period to learn the PC-MSN synaptic strength with Equation~(\ref{neoHeb}) by performing a period of replay. After that, the rat performs 100 test trials starting from each integer grid point in the maze (i.e., (1,1), (1,2), ...). Every test trial contains at most 6000 simulation steps. Before every 150 simulation steps (i.e., 3 s), the rat performs 1 s awake replay to evaluate the MSN activity along possible paths. Then, it samples a new direction from Equation~(\ref{behavPol}) and runs along this direction for next 100 simulation steps (i.e., 2 s). Once the rat enters a 1$\times$1 m circle enclosing the goal location, this test trial is completed and treated as a successful trial. 
	
	\textbf{Goal-changing experiment.} In the goal-changing experiment, the rat first completes the time course of the goal-fixed experiment. After test trials, the goal location is suddenly changed as shown in Figure \ref{setup3}. After goal changing, the rat is put at the new goal location for a while to learn the synaptic strength from place cells to the goal cell population by Hebbian-like learning $\mathbf{U}^{(n+1)} = \mathbf{U}^{(n)} + \alpha_3 (\mathbf{r}(\mathbf{x}_g)^{\rm T} \cdot 1  - \mathbf{U}^{(n)})$, so that $U_i$ will learn the form specified under Equation~(\ref{goalCell}). After that, the rat re-enters a brief rest period to update the PC-MSN synaptic strength by replay with the updated goal cell population. Then the rat performs another 100 test trials. 
	
	\textbf{Detour experiment.} In the detour experiment, the rat first completes the time course of the goal-fixed experiment. After test trials, two critical passages in the maze are closed as shown in Figure \ref{setup4}. After spatial layout changing, the rat reperforms 50 exploration trials to update the inter-PC synaptic strength. After that, the rat reperforms a period of replay to update the PC-MSN synaptic strength. Then the rat performs another 100 test trials. 
	
	\textbf{Shortcut experiment.} In the shortcut experiment, the rat first completes the time course of the detour experiment. After test trials, three walls in the right side of the detour maze are removed as shown in Figure \ref{setup5}. After the removement of walls, the rat once again randomly re-explores the maze for 50 trials and then reperforms a period of replay to update the PC-MSN synaptic strength. After that, the rat performs another 100 test trials. 
	
	\textbf{Experimental details.} In above experiments, the preferential locations of 2500 place cells are arranged on a regular grid of the maze without considering grid points that fall into areas under walls. The shortest path distance between pairs of preferential locations is computed by the Lee algorithm \cite{Lee1961}, based on breadth-first search (BFS) \cite{Cormen2009}. For Hebbian-like learning during exploration, $\sigma$ is 0.3 and $\alpha_1$ is 0.001. To simulate the dynamic of the CAN, $\Delta t$ is 1 ms, $h_0$ is 0, $\tau_r$ is 2 ms, $\tau_I$ is 0.5 s, $c_I$ is 10, and the global inhibitory strength added on the inter-PC synaptic strength is -0.3. For replay during the rest period, $\alpha_2$ is 0.01, $q$ is 0.1, $\tau_z$ is 0.5 s, $\xi$ is 0.3 and the duration is 60 s. The initial condition of the CAN during the rest period is a transient external input with amplitude 10 centered at the goal location for 10 ms. For replay during goal-directed navigation, the external input is persistent with amplitude 50, the radius $d$ is 0.5 m, $\beta$ is 10, and the duration is 1 s.
	
	\subsection{Experimental results}
	
	\begin{figure*}[h]
		\centering
		\subfigure[\label{exp1_a}]{\includegraphics[width=0.8in, height=0.8in]{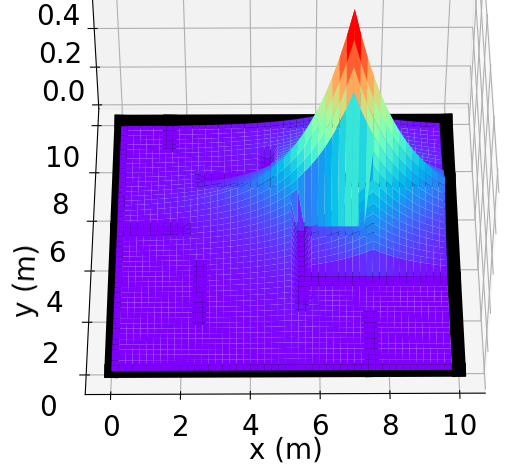}}
		\hfil
		\subfigure[\label{exp1_b}]{\includegraphics[width=1.0in, height=0.8in]{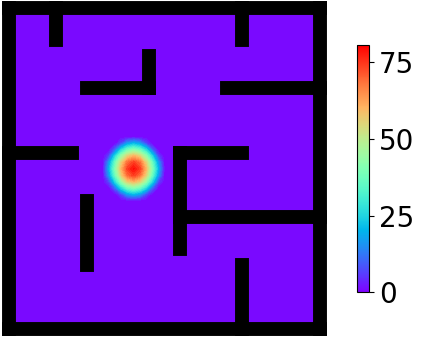}}
		\hfil
		\subfigure[\label{exp1_c}]{\includegraphics[width=0.8in, height=0.8in]{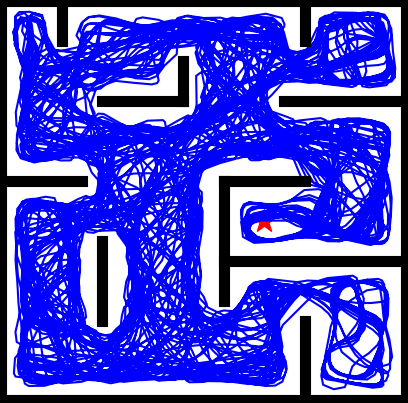}}
		\hfil
		\subfigure[\label{exp1_d}]{\includegraphics[width=0.9in, height=0.9in]{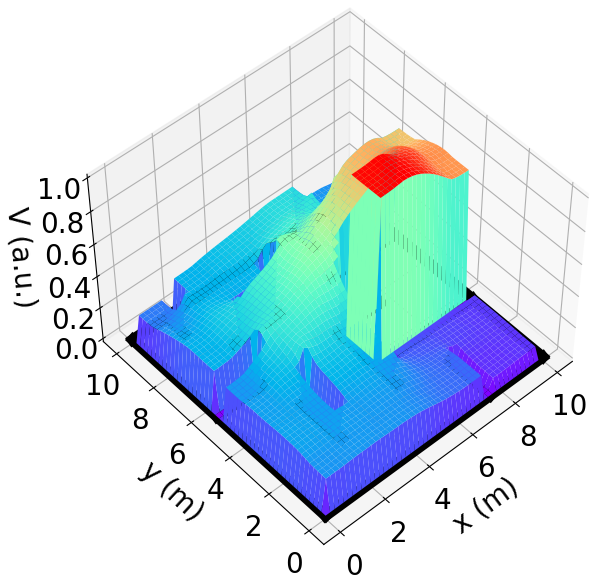}}
		\hfil
		\subfigure[\label{exp1_e}]{\includegraphics[width=0.8in, height=0.8in]{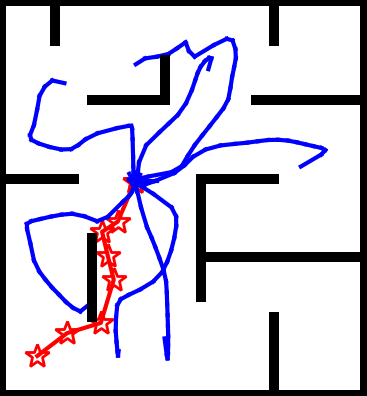}}
		\hfil
		\subfigure[\label{exp1_f}]{\includegraphics[width=0.8in, height=0.8in]{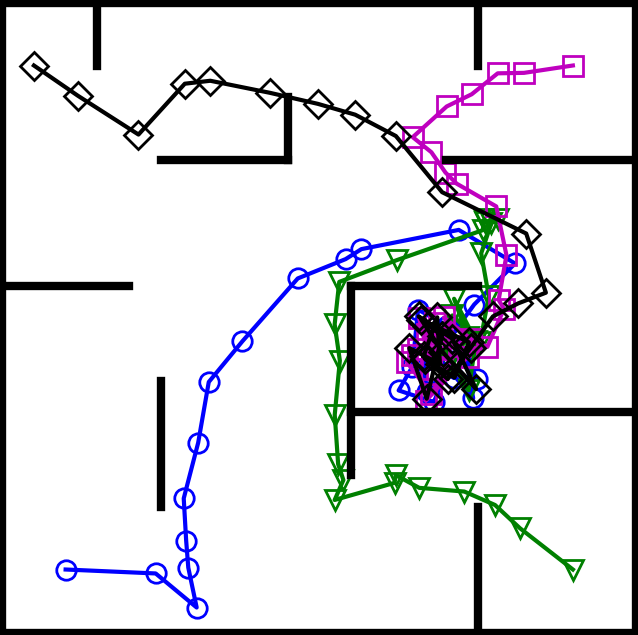}}
		\caption{Results for the goal-fixed experiment. (a) Inter-PC synaptic strength after exploration. (b) A stationary activity bump of the CAN under a strong external input. (c) The replay trajectory during the rest period. (d) The activity of the MSN population. (e) The replay sub-trajectories during planning (the red trajectory is the  rat's movement trajectory, and blue trajectories are replay sub-trajectories starting from the rat's location). (f) Exemplary movement trajectories during test trials.}
		\label{}
	\end{figure*}
	
	\textbf{Goal-fixed experiment.} Figure \ref{exp1_a} shows the strength of synapses projected from the place cell with preferential location (7, 6.4) to other place cells after exploration. The synaptic strength decays from the location (7, 6.4) along shortest paths to other locations, so the geodesic proximity between preferential locations of place cells is encoded by inter-PC synaptic strength. Figure \ref{exp1_b} shows a stationary activity bump centered at the rat's location under a strong external input with amplitude 100. After removing the external input during the rest period, the activity bump starts to deviate from its current position, and Figure \ref{exp1_c} shows the trajectory of the population vector of place cells. Interestingly, the trajectory can go around walls and well cover the whole maze. After replay during rest, the activity of the MSN, as a function of the location of the rat with a stationary activity bump, is shown in Figure \ref{exp1_d}. Interestingly, the activity of MSN will ramp up when the rat approaches the goal location from any initial locations, which conforms to a series of observations about MSN \cite{Meer2011, ONeal2022, Atallah2014, Sjulson2018, London2018, Howe2013}. Accordingly, the MSN activity encodes the geodesic proximity between each location and the goal location. Figure \ref{exp1_e} shows the sub-trajectories generated during a planning period in test trials. Strikingly, without any randomization mechanisms involved in the dynamic, the sub-trajectories can nearly uniformly explore each possible path to a considerable depth. This is especially beneficial if the MSN activity at a local area is either flat (uninformative) or with a wrong shape (misguiding). The success rate of test trials is 100\% and Figure \ref{exp1_f} shows several exemplary locomotion trajectories.
	
\begin{figure*}[h]
		\subfigure[\label{exp2_a}]{\includegraphics[width=0.8in, height=0.8in]{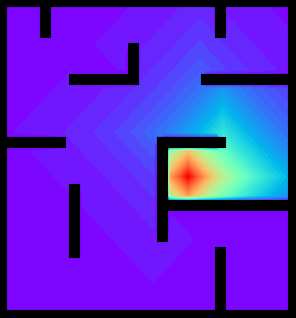}}
		\hfil
		\subfigure[\label{exp2_b}]{\includegraphics[width=0.8in, height=0.8in]{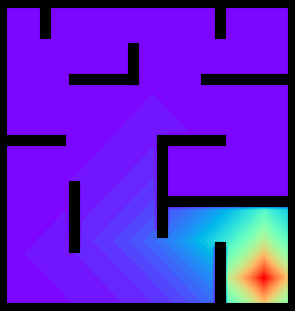}}
		\hfil
		\subfigure[\label{exp2_c}]{\includegraphics[width=0.9in, height=0.9in]{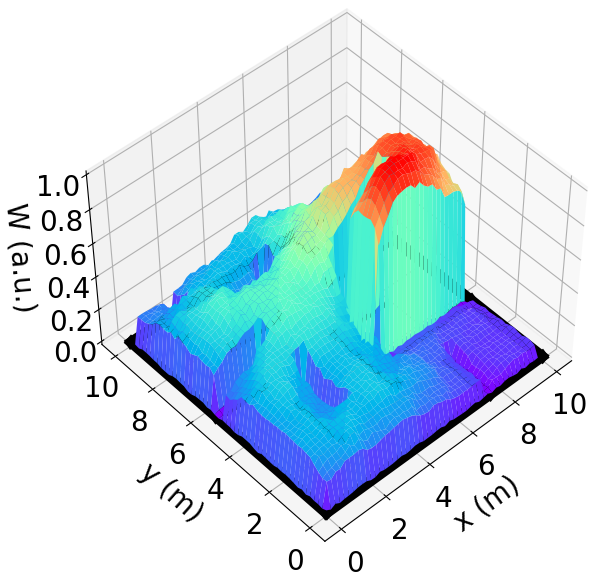}}
		\hfil
		\subfigure[\label{exp2_d}]{\includegraphics[width=0.9in, height=0.9in]{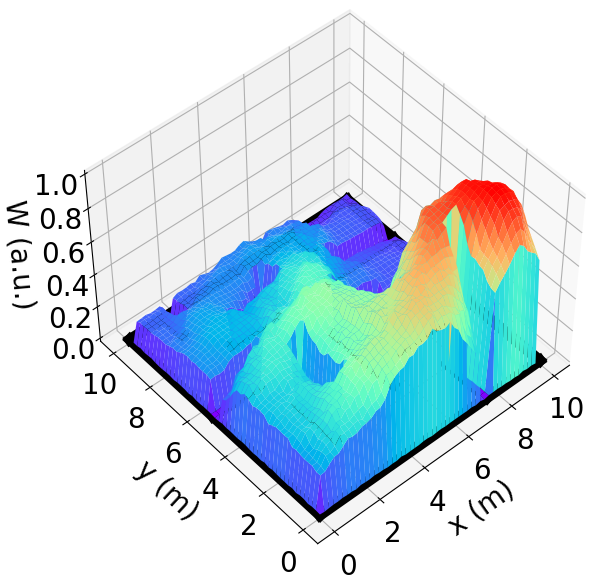}}
		\hfil
		\subfigure[\label{exp2_e}]{\includegraphics[width=0.9in, height=0.9in]{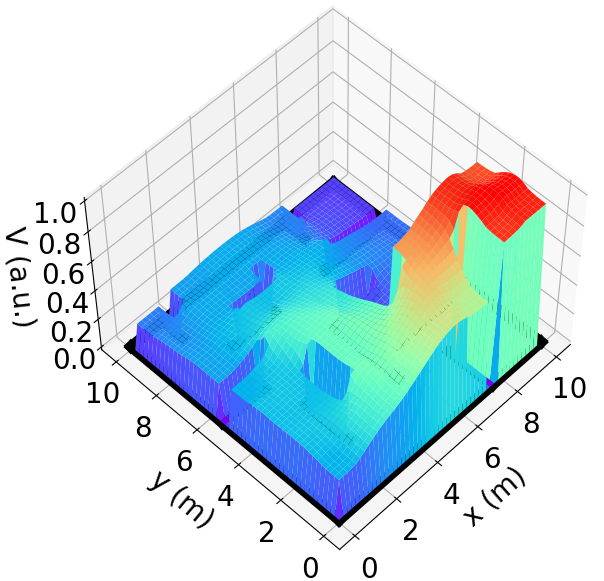}}
		\hfil
		\subfigure[\label{exp2_f}]{\includegraphics[width=0.8in, height=0.8in]{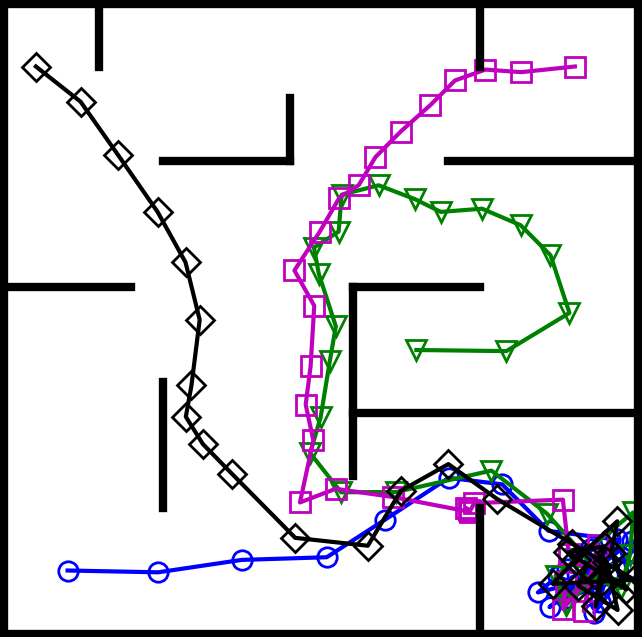}}
		\caption{Results for the goal-changing experiment. The place field of the goal cell population before (a) and after (b) goal changing. The PC-MSN synaptic strength before (c) and after (d) reperforming replay during rest. (e) The activity of the MSN population. (f) Exemplary movement trajectories during test trials.}
		\label{}
	\end{figure*}

	\textbf{Goal-changing experiment.} Figure \ref{exp2_a} and \ref{exp2_b} show the place fields of the goal cell population before and after goal changing, respectively. The place fields are constrained by walls and decay along paths away from the center. Before goal changing, the PC-MSN synaptic strength encodes the geodesic proximity between each location and the old goal location (Figure \ref{exp2_c}). After reperforming replay during rest, the PC-MSN synaptic strength now re-encodes the geodesic proximity between each location and the new goal location (Figure \ref{exp2_d}). As a result, the MSN activity ramps up along paths to the new goal location (Figure \ref{exp2_e}). The success rate of test trials after goal changing is still 100\% and Figure \ref{exp2_f} shows several exemplary locomotion trajectories.
	
\begin{figure*}[h]
		\subfigure[\label{exp3_a}]{\includegraphics[width=0.9in, height=0.9in]{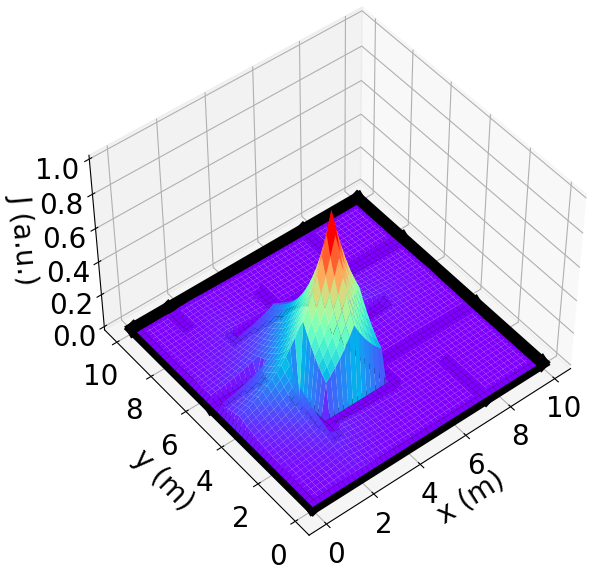}}
		\hfill
		\subfigure[\label{exp3_b}]{\includegraphics[width=0.8in, height=0.8in]{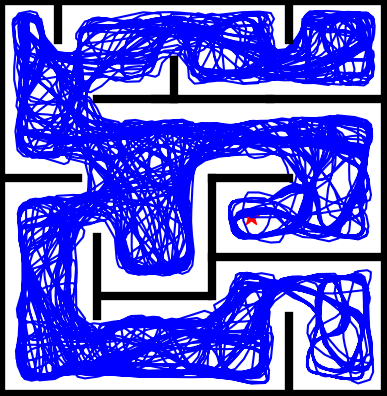}}
		\hfill
		\subfigure[\label{exp3_c}]{\includegraphics[width=0.9in, height=0.9in]{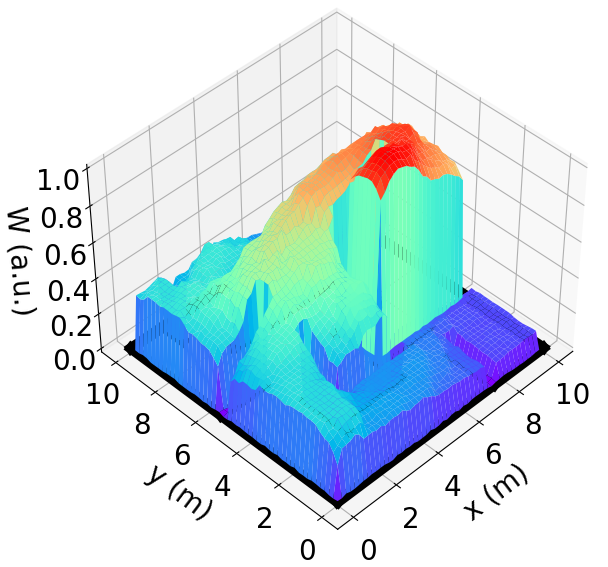}}
		\hfill
		\subfigure[\label{exp3_d}]{\includegraphics[width=0.9in, height=0.9in]{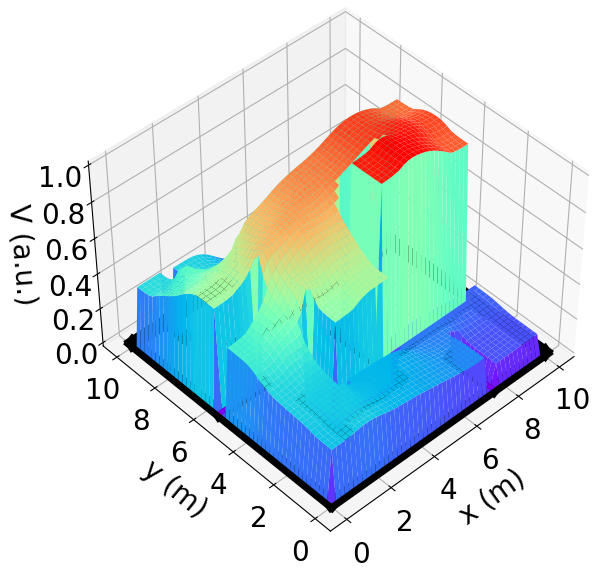}}
		\hfill
		\subfigure[\label{exp3_e}]{\includegraphics[width=0.8in, height=0.8in]{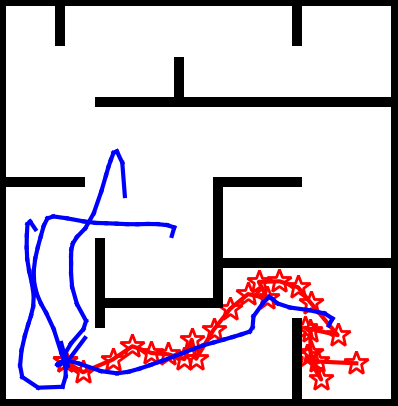}}
		\hfill
		\subfigure[\label{exp3_f}]{\includegraphics[width=0.8in, height=0.8in]{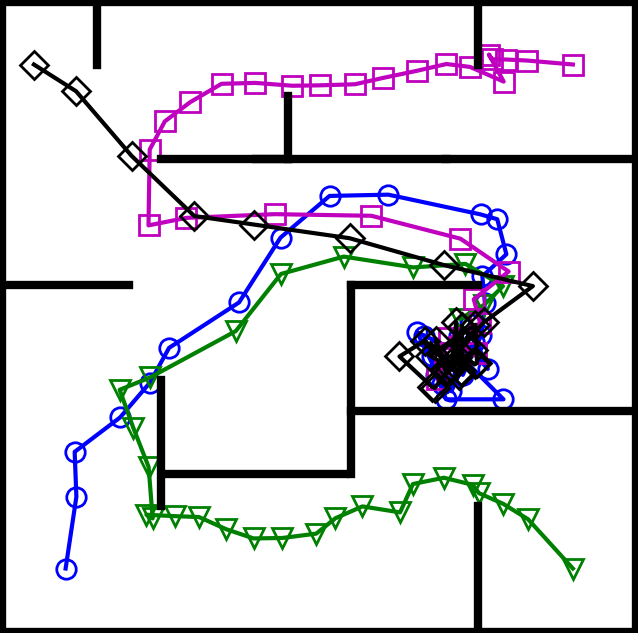}}
		\hfill
		\caption{Results for the detour experiment. (a) Inter-PC synaptic strength after re-exploration. (b) The replay trajectory after introducing new walls. (c) The PC-MSN synaptic strength after replay during rest. (d) The activity of the MSN population. (e) Replay sub-trajectories during planning. (f) Exemplary movement trajectories during test trials.}
		\label{}
	\end{figure*}

	\textbf{Detour experiment.} After re-exploration of the detour maze, the inter-PC synaptic strengths are updated and the strength of synapses projected from the place cell with preferential location (4, 4) to other place cells is shown in Figure \ref{exp3_a}. The inter-PC synaptic strength re-encodes the updated adjacency relation between locations in the detour maze. Under updated inter-PC synaptic strength, the replay trajectory during rest is blocked by new walls (Figure \ref{exp3_b}), signifying an adaptation to the new layout. After replay, both the PC-MSN synaptic strength and the MSN activity will ramp up along detours to the goal location (Figure \ref{exp3_c} and \ref{exp3_d}). Interestingly, the replay sub-trajectories during planning explore the area behind the new wall and evaluate the MSN activity there (Figure \ref{exp3_e}). As a result, the virtual rat achieves a 100\% success rate during test trials in the detour maze and Figure \ref{exp3_f} shows several exemplary movement trajectories. Starting from initial positions far away from the goal location, the rat can find a detour to the goal location. Such detour behavior resembles that observed in rodents \cite{Alvernhe2011, Tolman1930}, which has been considered a hallmark of cognition \cite{Tolman1948, Widloski2022}.
	
\begin{figure*}[h]
		\subfigure[\label{exp4_a}]{\includegraphics[width=0.9in, height=0.9in]{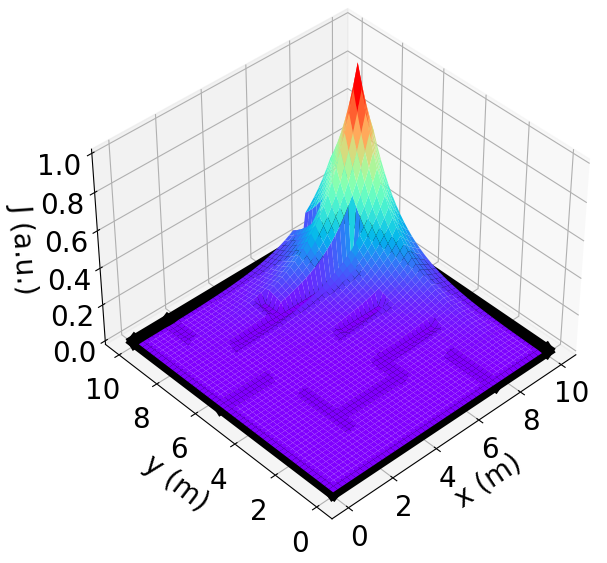}}
		\hfil
		\subfigure[\label{exp4_b}]{\includegraphics[width=0.8in, height=0.8in]{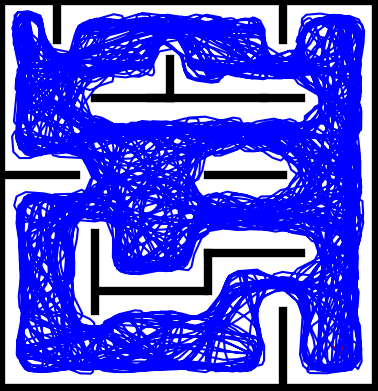}}
		\hfil
		\subfigure[\label{exp4_c}]{\includegraphics[width=0.9in, height=0.9in]{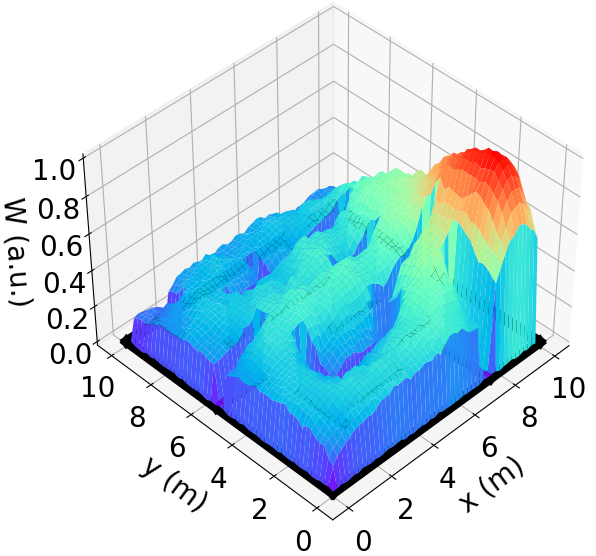}}
		\hfil
		\subfigure[\label{exp4_d}]{\includegraphics[width=0.9in, height=0.9in]{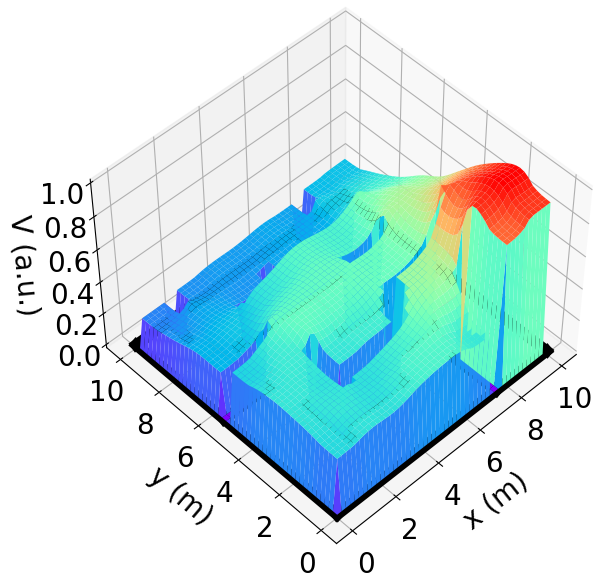}}
		\hfil
		\subfigure[\label{exp4_e}]{\includegraphics[width=0.8in, height=0.8in]{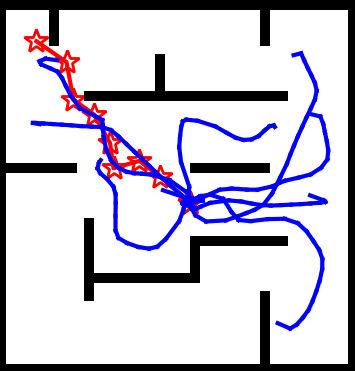}}
		\hfil
		\subfigure[\label{exp4_f}]{\includegraphics[width=0.8in, height=0.8in]{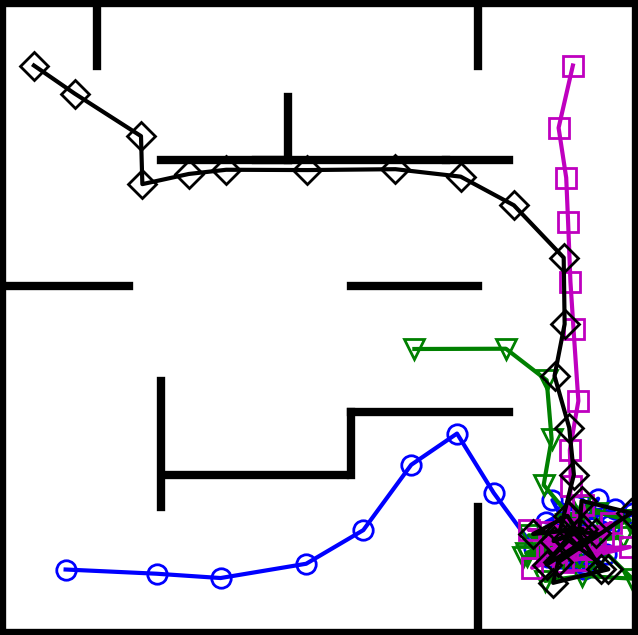}}
		\hfil
		\caption{Results for the shortcut experiment. (a) Inter-PC synaptic strength after re-exploring the shortcut maze. (b) The replay trajectory after removing walls. (c) The PC-MSN synaptic strength after replay during rest. (d) The activity of the MSN population. (e) Replay sub-trajectories during planning. (f) Exemplary movement trajectories during test trials.}
		\label{}
	\end{figure*}

	\textbf{Shortcut experiment.} After re-exploring the shortcut maze, the strength of synapses projected from the place cell with preferential location (9, 9) to other place cells decays along newly available shortcuts (Figure \ref{exp4_a}). As a result, the replay trajectory during rest will freely traverse these shortcuts (Figure \ref{exp4_b}). After replay, both the PC-MSN synaptic strength and the MSN activity ramp up along newly available shortcuts (Figure \ref{exp4_c} and \ref{exp4_d}). The replay  sub-trajectories during planning can lookahead along newly available shortcuts (Figure \ref{exp4_e}). The success rate during test trials in the shortcut maze is 100\% and Figure \ref{exp4_f} shows several exemplary locomotion trajectories, all following the shortcut to the goal location.
	
	\textbf{The influence of parameters on performance.} To measure the performance of the virtual rat during test trials, we define the normalized latency of a successful trial as the time taken to reach the goal location divided by the shortest path distance between the initial position and the goal location. The normalized latency eliminates the influence of different initial positions on the time required to reach the goal. It measures the average time taken to approach the goal location per meter hence it characterizes the intrinsic efficiency of the navigation independent of initial positions. 
	
\begin{figure*}[h]
		\subfigure[\label{exp5_a}]{\includegraphics[width=1.15in, height=0.85in]{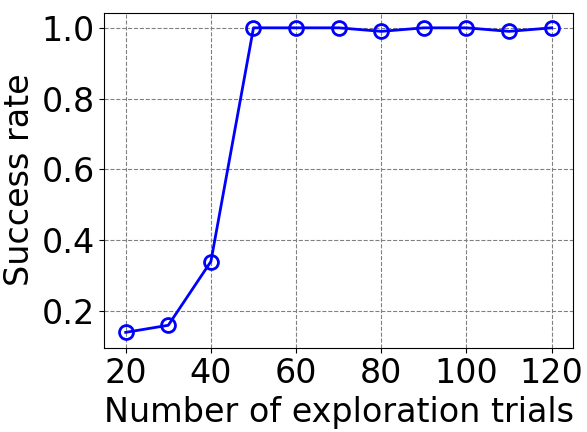}}
		\hfil
		\subfigure[\label{exp5_b}]{\includegraphics[width=1.15in, height=0.9in]{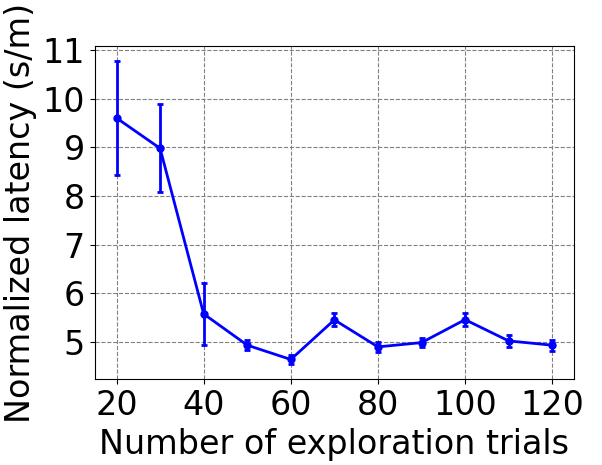}}
		\hfil
		\subfigure[\label{exp5_c}]{\includegraphics[width=1.15in, height=0.85in]{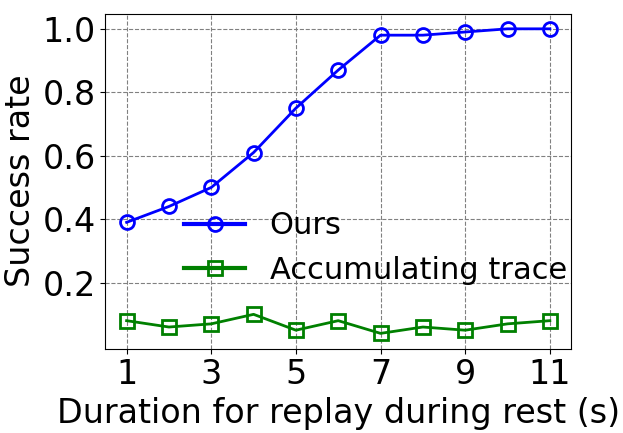}}
		\hfil
		\subfigure[\label{exp5_d}]{\includegraphics[width=1.15in, height=0.9in]{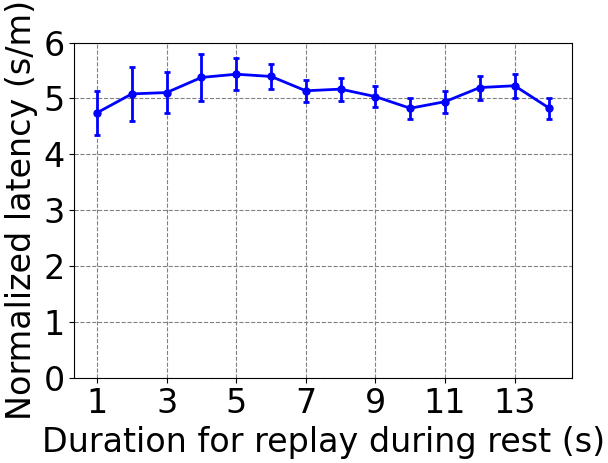}}
		\hfil
		\subfigure[\label{exp5_e}]{\includegraphics[width=1.2in, height=0.85in]{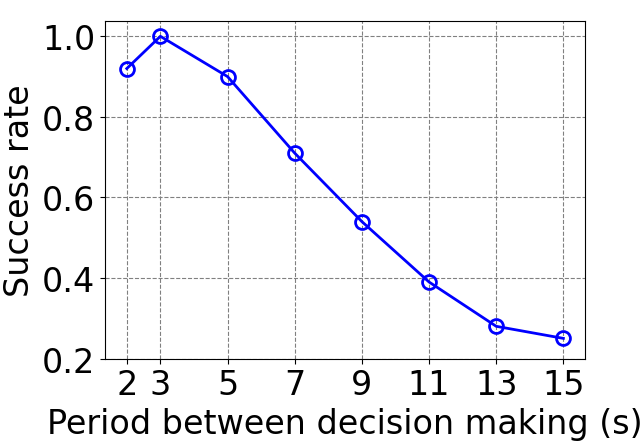}}
		\hfil
		\subfigure[\label{exp5_f}]{\includegraphics[width=1.15in, height=0.9in]{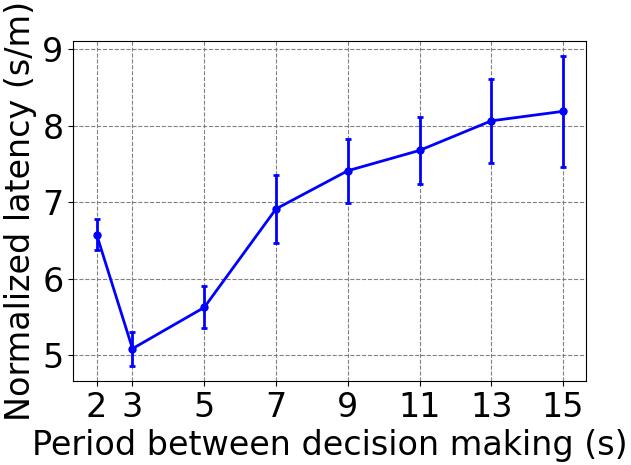}}
		\hfil
		\subfigure[\label{exp5_g}]{\includegraphics[width=1.15in, height=0.85in]{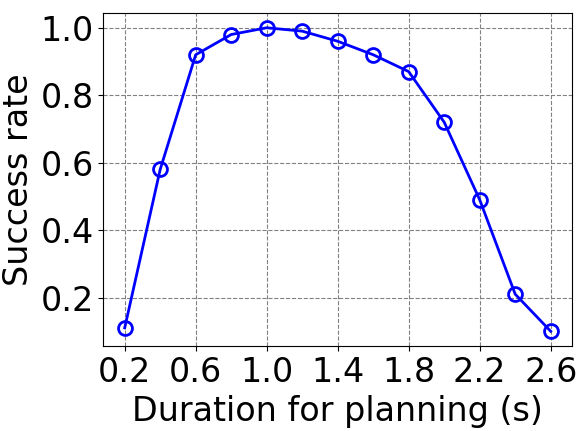}}
		\hfil
		\subfigure[\label{exp5_h}]{\includegraphics[width=1.15in, height=0.9in]{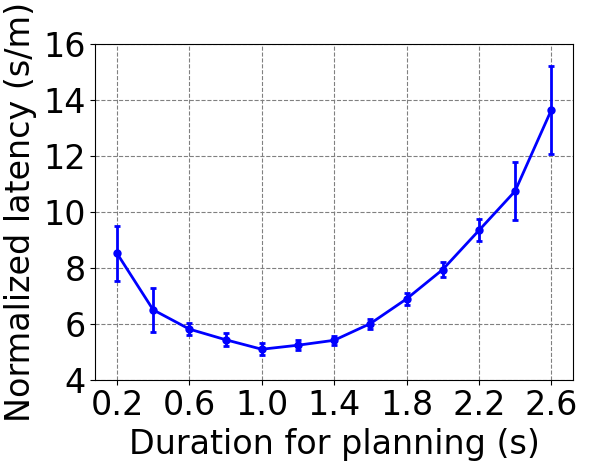}}
		\hfil
		\subfigure[\label{exp5_i}]{\includegraphics[width=1.15in, height=0.85in]{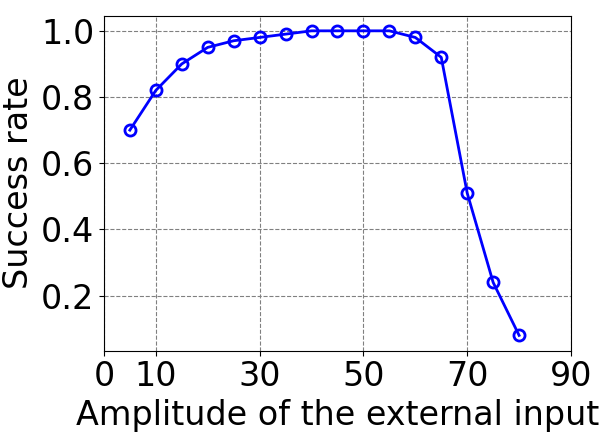}}
		\hfil
		\subfigure[\label{exp5_j}]{\includegraphics[width=1.15in, height=0.9in]{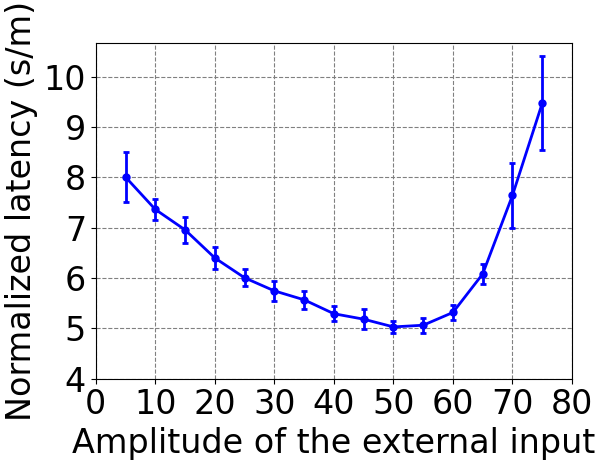}}
		\hfil
		\caption{The influence of parameters on performance. (a) and (b) are the influence of the number of exploration epochs. (c) and (d) are the influence of duration for replay during rest. (e) and (f) are the influence of the period between decision making. (g) and (h) are the influence of duration for planning. (i) and (j) are the influence of the external input.}
		\label{}
	\end{figure*}

	To show the influence of parameters on performance, the goal-fixed experiment is repeated under varying values of a considered parameter while keeping other parameters as default values. As Figure \ref{exp5_a} shows, the success rate of test trials improves when the number of trials for exploration increases but the success rate nearly keeps at 1 as long as at least 50 exploration trials are performed. As shown in Figure \ref{exp5_b}, the normalized latency reduces with an increasing number of exploration trials and fluctuates around 5 s/m after 50 exploration trials. With an inadequate number of exploration trials, the spatial experiences of the virtual rat can’t fully cover the maze. As a result, the synaptic strength between place cells with preferential locations at those under-explored areas are weak, which prevents the replay during rest to fully explore the maze and impairs the learning of PC-MSN synaptic strength consequently. 
	
	As Figure \ref{exp5_c} shows, the success rate increases with longer duration for replay during rest and saturates at nearly 1 as long as at least 7 seconds are allowed. Although the MSN activity after only 7 seconds of replay has only a rough and unsmooth trend to ramp up (not shown),  the planning trajectories will lookahead over a long distance, enough to overcome the local unsmoothness and utilize the long-range trend to find the correct direction to the goal location. Such rapid learning is also observed in rodent experiments \cite{Rosenberg2021} and it might be very important for the survival of rodents in a changing environment. In contrast, the three-factor rule using the accumulating trace update $\dot{z}_i(t) = - z_i(t)/\tau_z + r_i(t) V(t)$ \cite{Gerstners2018, Izhikevich2007} fails to improve the success rate with an increasing duration for replay. The reason for the inferior performance is that the accumulating trace overly strengthens the PC-MSN synapses of frequently activated place cells even if their firing field centers are far away from the goal location. Such frequency-dependence leads to local peaks of the MSN activity that might attract the rat at places without rewards. As shown in Figure \ref{exp5_d}, the normalized latency of successful trials fluctuates around 5 s/m and shows little dependence on the duration for replay during rest. 
	
	As shown by Figure \ref{exp5_e} and \ref{exp5_f}, the success rate generally decreases and the normalized latency generally increases when the period between decision making is prolonged, because infrequent decision making fails to timely re-adjust the movement direction that has became suboptimal over a distance of locomotion. In contrast, too frequent decision making (e.g., 2 s) also harms the performance because the overhead of planning (e.g., 1 s/2 s = 50\%) increases so that the time used for locomotion is reduced, which slowdowns the progress toward the goal location.
	
	As Figure \ref{exp5_g} and \ref{exp5_h} show, when the duration for planning is shorter than 1 s, the performance improves with a longer duration for planning due to the ability to evaluate more directions. In contrast, when the duration for planning is longer than 1 s, the performance deteriorates due to the increasing overhead of planning and the shortened time for locomotion. As  Figure \ref{exp5_i} and \ref{exp5_j} show, when the amplitude of the external input is smaller than 50, the performance improves with a larger external input, due to a reduced length of sub-trajectories leading to an increasing number of sub-trajectories during planning. However, when the amplitude of the external input is larger than 50, the performance collapses because it becomes more difficult to deviate the activity bump so that the number of sub-trajectories reduces rapidly.
	
	\section{Discussion}
	This paper proposed a computational model that uses a continuous attractor network (CAN) to generate layout-conforming replay for achieving reward-based learning and planning, two key functions that support flexible navigation in a dynamically changing maze. We have shown that these functions are achieved by a single CAN under a different strength of external inputs. Reward-based learning is implemented by three-factor learning of PC-MSN synaptic strength during replay under vanishing external inputs. Planning is achieved by evaluating the MSN activity along lookahead trajectories during replay under non-zero external inputs. Combining these two functions, this paper sheds a light on how the learning of flexible navigation in a maze is achieved by the activity of an ensemble of interacting place cells. Besides, our model is beneficial to the design of neuroscience-inspired artificial intelligence \cite{Hassabis2017}. We have shown that incorporating state-of-the-art neuroscientific insights about learning, memory and motivation into the design loop of an artificial agent is important toward developing artificial intelligence that matches the performance of animals.
	
	Many works use attractor states of a CAN to model the population activity of head-direction cells \cite{Skaggs1995, Blair1996, Zhang1996} or place cells \cite{Tsodyks1999, Samsonovich1997, McNaughton2006, Battaglia1998, Stringer2002}. In these models, the synaptic strength is assumed to be a negative exponential function of the Euclidean distance between preferential angles or locations of two cells, possibly with periodic boundary conditions. This assumption is valid for head-direction cells in a ring attractor space or place cells in a barrier-free plane attractor space but it ignores the influence of barriers in non-Euclidean spaces, such as a maze. 
	
	Next, we discuss previous computational models for the generation of replay with a CAN. Two computational models \cite{Azizi2013, Hopfield2009} use integrate-and-fire cells with spike-frequency adaptation (SFA) to simulate drifts of the activity bump. In these models, the SFA is modeled by an adaptive inhibitory current fed into each place cell, which is increased after every spike of a place cell. Despite functionally similar to the feedback inhibition in our model, the adaptive inhibitory current is used to model the dynamic blockage of Ca$^{2+}$-dependent K$^{+}$ channels \cite{Hopfield2009, Faber2003, Shao1999} rather than interactions between place cells and hippocampal GABAergic interneurons. One model \cite{Itskov2011} and another model \cite{Romani2015} use adaptive thresholds and short-term depression, respectively, to generate drifts of the activity bump. However, the generation of hippocampal replay requires interactions between place cells and inhibitory interneurons \cite{Schlingloff2014, Stark2014, Buzsaki2015}, which is not captured by these two models. Other models \cite{Spalla2021, Zhang1996} generate bump drifts by introducing an asymmetric component into the inter-PC synaptic strength. However, it remains unclear how such a systematic perturbation of inter-PC synaptic strength can be implemented in the hippocampus. Besides, the above models pre-configure the inter-PC synaptic strength as a function decaying with the Euclidean distance between preferential locations, so the bump drifts generated by these models will violate the constraints imposed by walls of a maze. Among the aforementioned models, some of them \cite{Azizi2013, Hopfield2009, Spalla2021, Romani2015} consider multiple possible mapping from place cells to place field centers, called multi-chart, to model a global remapping of place cells across different environments \cite{Alme2014}. However, it's observed that the place field centers are largely stable under different layout of a maze \cite{Widloski2022, Alvernhe2011}. Accordingly, we keep the mapping of place cells consistent across different layout.
	
	The early works \cite{Muller1996a, Muller1991} first proposed that the inter-PC synaptic strength should decay with the Euclidean distance between their firing field centers due to the long-term potentiation \cite{Isaac2009}. In these works \cite{Muller1996a, Muller1991}, the hippocampus is modeled as a weighted graph where place cells are nodes and synapses are edges with synaptic resistance (reciprocal synaptic strength) as edge weights. To navigate to the goal location, this model uses the Dijkstra's algorithm \cite{Cormen2009} to search a shortest path in the graph from the currently activated place cell to the place cell activated at the goal location. However, it remains unclear how such complex path search computations are implemented in rodent brains. Our model has shown that the learning of PC-MSN synaptic strength by replay during rest encodes the geodesic proximity between each place field center and the goal location, which is read out subsequently during awake replay to plan a path toward the goal location.
	
	If the firing rate vector of place cells is interpreted as a feature vector of the rat's location, then the activity of MSN can be interpreted as a linear value function approximation \cite{Sutton2018} of the rat's location. Such an interpretation relates our model to a line of spatial navigation models based on reinforcement learning (RL). In one model \cite{Brown1995}, the synaptic strengths from PC to motor neurons are learned using policy gradient like reinforcement learning (RL) \cite{Williams1992, Sutton1999} to build correct place-response associations by trial-and-errors. Another line of models \cite{Foster2000, Gustafson2011, Banino2018} use the firing rates of place cells as the set of basis functions for representing the policy and the value function, which are trained by the actor-critic algorithm \cite{Sutton1998} during interactions with the environment. Unfortunately, navigation behavior learned with such model-free RL algorithms  is inflexible to changes of the goal location or the spatial layout.
	
	Next, we discuss computational models of spatial navigation based on model-based RL. Inspired from the Dyna-Q algorithm \cite{Sutton1990}, a computational model \cite{Johnson2005} learns a Q-value function during offline replay of a network of place cells. This model interprets the inter-PC synaptic strength matrix as a state transition matrix that sequentially activates a place cell with a maximal synaptic strength with the currently activated place cell. Such one-by-one replay of place cells with binary activity can hardly capture the bump-like activity during hippocampal replay \cite{Widloski2022,Stella2019,Pfeiffer2013}. Besides, the Q-value function is learned by the biologically unrealistic one-step TD(0) algorithm (i.e., Q-learning) without using a chemical trace. The SR-Dyna model \cite{Russek2017, Momennejad2020a} learns a successor representation (SR) matrix by TD(0) during one-by-one replay of binary place cells. After replay, the Q-value function is computed simply by a dot product of the SR matrix and the reward function. However, the neural substrate of the SR remains unclear and controversial \cite{Russek2017, Stachenfeld2017}. 
	
	Next, we discuss computational models of spatial navigation without using a value representation. Two studies \cite{Gerstner1997, Blum1996} proposed that the long-term potentiation (LTP) of inter-PC synapses during exploration slightly shifts the location encoded by the population vector away from the rat's actual location. Such shifts are used as a vector field tending toward the goal location to navigate the rat to the goal. Our model similarly uses a shift of the population vector during awake replay to define the directional vector of each replay sub-trajectory. However, our model does not construct a static vector field toward the goal but relies on online planning with respect to the MSN activity to choose a direction to follow. Another model \cite{Burgess1996} directly uses the place field of a putative goal cell to navigate to the goal by moving along the gradient direction of the place field. However, the gradient of the place field will vanish at locations far away from the goal location due to the limited range of the goal cell's place field. In our model, the activity of MSN will ramp up even at locations far away from the goal location. Furthermore, planning with respect to the MSN activity can avoid wrong gradient directions due to local unsmoothness of the MSN activity. 
	
	The work \cite{Ponulak2013} uses a similar CAN as in \cite{Hopfield2009} to generate a wavefront propagation of activity on the sheet of place cells after visiting the goal location. With anti-STDP, the inter-PC synaptic strength will be biased toward the goal location after wavefront propagation, which can be interpreted as a synaptic vector field to guide the navigation toward the goal. However, the activity propagation during hippocampal replay is sequential and follows particular directions \cite{Pfeiffer2017} rather than propagating over all directions as in the above model. Another model \cite{Gonner2017} uses a learned goal location representation of place cells in dentate gyrus (DG) as the input of place cells in CA3 to move the activity bump toward the goal location. The end point of the moving bump is used for vector-based navigation. However, the bump trajectories generated by this model always converge to the goal location, falling short of explaining the diversity of directions and end points of replay trajectories.
	
	Next, we discuss limitations of our model. Since our model encodes the goal location information by the MSN activity, our modeling of place fields does not capture some weak off-center goal-related activity observed in \cite{Hok2007}. Besides, the ubiquitous existence of such goal-related activity of place cells remains controversial and it might introduce ambiguity to the coding of place cells \cite{Poucet2017}. Strictly speaking, the most detailed model of symmetric STDP \cite{Mishra2016} should consider spike trains of pre- and post-synaptic neurons. Since the pre- and post-synaptic spike trains of a pair of place cells can be modeled as independent Poisson processes before learning the inter-PC synaptic strength, the synaptic update under the rate-based Hebbian-like rule equals the average synaptic updates under symmetric STDP using spike trains \cite{Kempter1999}. Our model requires the exploration performed by the virtual rat fully covers the whole maze. However, animals can navigate in unexplored areas based on landmarks. Our model of the feedback inhibition in the CAN is a simplified abstraction of the interaction between place cells and interneurons without modeling mutual interactions among interneurons \cite{Schlingloff2014, Stark2014, Buzsaki2015}. Our model of the MSNs does not capture observations that the activity of some MSN assemblies encodes other behavioral information, such as locomotion initialization \cite{Kravitz2010, Freeze2013} or speed \cite{Fobbs2020, Kim2014}.
	
	\bibliography{main}
	\appendix
	
	\section{Appendix}
	\subsection{Pretraining details}
	\label{pretrain}
	In order to train the policy that turns $\theta^\circ$ then runs forward, we use the following reward function,
	\begin{equation}
		r = {vel}_g + 0.1 \vec{h}_g \cdot \vec{h}_r + 0.1 \vec{n}_z \cdot \vec{n}_r.
	\end{equation}
	Here ${vel}_g$ is the velocity of the rat along the targeted direction. $\vec{h}_g$ is the unit vector along the targeted direction. $\vec{h}_r$ is the unit vector along the body direction of the rat. $\vec{n}_z$ is the unit vector along the $z$-axis in the global coordinate frame. $\vec{n}_r$ is the unit vector along the $z$-axis in the local coordinate frame of the rat. To obtain high rewards, the rat requires to turn into the targeted direction and run as fast as possible while keeping $\vec{n}_r$ upward to avoid falling down. The TD3 algorithm trains each policy network from scratch to maximize the accumulated rewards collected from a behavior trajectory. Training each policy takes several hours on a GPU server.
	
	\subsection{Hyperparameters of the TD3 algorithm}
	\label{hyperparas}
	\begin{table}[h]
		\caption{Hyperparameters and their values}
		\label{para-table}
		\centering
		\begin{tabular}{ll}
			\toprule
			Hyperparameter &  Value  \\
			\midrule
			Replay buffer size &  100000 \\
			Learning rate & 0.0001 \\
			Exploration coefficient & 0.2 \\
			Discount factor & 0.98 \\
			Mini-batch size & 256 \\
			Number of layers & 5 \\
			Target net update rate & 0.01 \\
			Noise standard deviation & 0.2 \\
			Noise clip coefficient & 0.5 \\
			\bottomrule
		\end{tabular}
	\end{table}
	
	\end{document}